\documentclass[11pt]{article}
\usepackage{amsmath,amsfonts}
\usepackage{verbatim}
\usepackage{relsize}
\usepackage[usenames]{color}
\normalbaselines
\newtheorem{teor}{Theorem}
\newtheorem{prop}[teor]{Proposition}
\newtheorem{lema}[teor]{Lemma}
\newtheorem{coro}[teor]{Corollary}
\newtheorem{rem}[teor]{Remark}

\newtheorem{defi}[teor]{Definition}

\newenvironment{demo}{\rm \trivlist \item[\hskip \labelsep{\it
      Proof}.]}{\nopagebreak \hfill $\square$ \endtrivlist}

\title{From holography to the geometry of the spacetime}

\author{Jos\'e A. S. Pelegr\'in \\[6mm]
Departamento de Geometr\'\i a y Topolog\'\i a, \\ [0.5mm]
Universidad de Granada, 18071 Granada, Spain \\ E-mail\textup{:
\texttt{jpelegrin@ugr.es}} \\[3mm]}

\date{}

\begin{document}

\maketitle

\thispagestyle{empty}

\begin{abstract}
We show the geometric consequences that the holographic principle has on the spacetime. Namely, we prove that complete spacelike hypersurfaces in a spacetime that satisfies the holographic principle are non-parabolic. This has important consequences on the behaviour of the Brownian motion in these spacelike hypersurfaces, as well as provides a method for finding examples of universes that lead to a violation of this principle.
\end{abstract}
\vspace*{5mm}

\noindent \textbf{MSC 2010:} 53C80, 58J05, 53C50

\noindent  \textbf{Keywords:} Holographic principle, transience, spacelike hypersurface.

\section{Introduction}

The recent revolution in string theory and black holes theory have changed the way we understand the nature of the universe. Among the new unexpected results that have been obtained, one of utmost importance is the holographic principle, a mathematical conjecture that describes the relation between the spacetime geometry and the number of quantum states of matter. Namely, the degrees of freedom of a spatial region reside not in the interior but on its boundary. This fact reduces the complexity of the physical systems and, supported by the AdS/CFT duality \cite{W}, it has been argued that a fundamental theory that aims to unify gravity and quantum mechanics should include this counterintuitive result, i.e., it should be a holographic theory \cite{B2}. Our aim in this article is to obtain some geometric consequences derived from the assumption of this holographic principle in a relativistic universe.

Historically, the first step towards the holographic principle was due to Bekenstein \cite{Be} in 1981, when guided by the second law of thermodynamics he obtained that the entropy in a spherically symmetric spatial region of the spacetime $S(V)$ satisfied

$$S(V) \leq \frac{\mathrm{Area}(\partial V)}{4},$$

\noindent being $\mathrm{Area}(\partial V)$ the area of the boundary of $V$ in Planck units. This bound is saturated ,i.e., equality holds, by the
Bekenstein-Hawking entropy associated with a black hole horizon. As a result, no
stable spherical system can have a higher entropy than a black hole of equal size. 

Thermodinamical entropy $S$ has a statistical interpretation as the logarith of the number of independent quantum states compatible with the macroscopic state of a physical system \cite{B1}. Therefore, the dimension of the system's Hilbert space is given by $dim(\mathcal{H}) = e^S$. This enabled 't Hooft \cite{tH} and Susskind \cite{S} to strengthen the interpretation of Bekenstein's result, stating that the number of independent degrees of freedom contained in a spatial volume $\textit{N}_{dof}(V)$ (which is defined as the logarithm of the dimension of the system's Hilbert space) satisfies

$$\textit{N}_{dof}(V) \leq \frac{\mathrm{Area}(\partial V)}{4}.$$

Thus, a physical system can be completely determined by data stored on its boundary without exceeding a density of one bit per Planck area \cite{B2}.

In this article, by means of some potential theory results we are able to obtain geometric properties of the spacelike hypersurfaces immersed in a spacetime (i.e., a connected time-orientable Lorentzian manifold endowed with a time-orientation) of arbitrary dimension where the holographic principle holds. Given an $n$-dimensional manifold $M$, an immersion $\psi: M \longrightarrow \overline{M}$ in an $(n+1)$-dimensional spacetime $(\overline{M}, \overline{g})$ is called a spacelike hypersurface if $g=\psi^{*}\overline{g}$ is a Riemannian metric on $M$ \cite{O'N}. Roughly speaking, spacelike hypersurfaces describe the physical universe that some observers can measure in a given instant of their proper time. These geometric objects are of great importance in General Relativity both from a physical and mathematical perspective. Among other reasons, we recall that the initial value problem for the Einstein's field equation is formulated in terms of a spacelike hypersurface \cite{Ri} and even in causality theory the existence of a certain spacelike hypersurface can determine the causal properties of the spacetime \cite{Ger}.

Hence, once we show the restrictions that the holographic principle imposes on the geometry of the spacelike hypersurfaces, the search of models that lead to a violation of this principle is reduced to finding spacetimes with spacelike hypersurfaces where these geometric restrictions do not hold. In particular, we obtain that this holographic principle is violated in the so called spatially parabolic spacetimes, i.e., those that admit a parabolic spacelike hypersurface \cite{RRS}. Note that spatially parabolic spacetimes are a generalization of spatially closed models (i.e., spacetimes that admit a compact spacelike hypersurface). In fact, it was proved in \cite{B1} that for a certain entropy distribution spatially closed spacetimes lead to a violation of the holographic principle. 

This article is organized as follows. In Section \ref{sepr} we fix the formulation of the holographic principle that we will use throughout this paper, namely, the spacelike entropy bound. Moreover, we will also describe some properties of the Brownian motion on a Riemannian manifold and how it is affected by the manifold's geometry. In Section \ref{s2} we obtain our main results concerning the influence of the holographic principle on the geometry of complete spacelike hypersurfaces in arbitrary spacetimes. Finally, we devote Section \ref{segrw} to particularize our previous results for Generalized Robertson-Walker spacetimes that obey the holographic principle and obtain several results for the behaviour of the Brownian motion in spacelike hypersurfaces in these ambient spacetimes.

\section{Preliminaries}
\label{sepr}

In this section we will fix the formulation of the holographic principle that we will later use in order to obtain our results as well as recall some aspects of the study of the Brownian motion on Riemannian manifolds that will appear along this article.

\subsection{The spacelike entropy bound}
\label{seseb}

Since Bekenstein's bound only holds for spherically symmetric weakly gravitating matter systems in assymptotically flat spacetimes \cite{Be}, we will use the following generalization, which is known as the spacelike entropy bound (see \cite{B1} and \cite{tH}).

\begin{defi}
The spacelike entropy bound is said to hold in a spacetime $\overline{M}$ if for every spacelike hypersurface $\psi: M \longrightarrow \overline{M}$ in $\overline{M}$ the entropy of all matter systems contained in any geodesic ball $B_R \subset M$, $S(B_R)$, satisfies

$$S(B_R) \leq \frac{\mathrm{Area}(\partial B_R)}{4},$$

\noindent where $\mathrm{Area}(\partial B_R)$ is the area of the boundary of $B_R$.
\end{defi} 

This formulation of the holographic principle has been criticized (although some of the arguments against the spacelike entropy bound are weakened by the results obtained in \cite{Te}), existing several alternative ones. For instance, Fischler and Susskind \cite{FS} proposed to constrain only the entropy passing through the past light cone. Later, continuing with these ideas of trying to establish a lightlike formulation for the holographic principle, Bousso \cite{B2} proposed a covariant entropy bound in terms of light sheets. It has also been stated in terms of the cosmological apparent horizon \cite{BR} as well as taking other cosmological considerations into account (see \cite{BV} and \cite{EL}). 

However, every formulation of the holographic principle has received criticism. In fact, the controversy generated is mostly due to the examples of universes that violate each possible formulation (see \cite{B1}, \cite{BV} and \cite{KL}, for instance). As we will show in this article, the validity of the holographic principle has influence on the geometry of the spacetime. Therefore, the search of counterexamples to each formulation is reduced to finding spacetimes that do not satisfy the associated geometric restrictions. 

\subsection{Brownian motion on a Riemannian manifold}
\label{sebm}

The irregular movement of microscopic particles suspended in a liquid has been known since 1828, when Brown studied the erratic motion of pollen grains on water. However, it was not until 1905 that Einstein explained this effect, obtaining that it satisfies a diffusion equation. Later, Wiener \cite{Wi} was able to construct a rigorous continuous model of the Brownian motion.

Thanks to Wiener's model, we can study the behaviour of a particle with a Brownian motion on an arbitrary Riemannian manifold. In fact, the easiest way to construct Brownian motion on a Riemannian manifold $(M, g)$ is by means of the heat kernel $p(x, y, t)$, a function on $M \times M \times (0, +\infty)$ which is the smallest positive fundamental solution of the heat equation

$$ \frac{\partial p}{\partial t} - \frac{1}{2} \Delta p = 0 \ \ \ \textnormal{on} \ M,$$

\noindent in the variables $(x, t)$ (the point $y$ is fixed), with initial data

$$ p(\cdot, y, t) \rightarrow \delta_y \ \ \ \textnormal{for} \ t \rightarrow 0^+.$$

Indeed, Dodziuk \cite{Do} proved the existence of the heat kernel on any arbitrary Riemannian manifold regardless of geodesic completeness. The properties of the heat kernel allow us to construct a (sub)Markov process $X_t$ on $M$ with transition density $p$ (see \cite{CZ}), which is a diffusion process known as Brownian motion on $M$. If we denote by $\mathbb{P}_x$ corresponding measure in the space of paths emanating from $x$, we can define the notions of recurrence and trancience as in \cite[Def. 2.1]{G}.

\begin{defi}
Brownian motion $X_t$ on a Riemannian manifold $M$ is recurrent if, for any non-empty open set $\Omega$ and for any point $x \in M$,

$$\mathbb{P}_x \{\textnormal{there is a sequence} \ t_k \rightarrow + \infty \ \textnormal{such that} \ X_{t_k} \in \Omega \} = 1.$$

\noindent Otherwise $X_t$ is transient.
\end{defi}

Moreover, by means of geometric analysis' techniques, it has been proved that the recurrence of the Brownian motion on a manifold is equivalent to that manifold being parabolic \cite{G}. Therefore, the Brownian motion will be transient on non-parabolic Riemannian manifolds. We recall that a complete Riemannian manifold $M$ is parabolic if the only bounded from below, superharmonic $C^2$-functions on $M$ are the constants. Hence, the development of potential theory and the study of parabolicity has greatly contributed to the better understanding of the Brownian motion's recurrence \cite{LT}.

In fact, whereas in the Euclidean space $\mathbb{R}^n$ the behaviour of the Brownian motion only depends on the dimension (being recurrent if $n \leq 2$ and transient otherwise), in an arbitrary Riemannian manifold recurrence is related to some geometric properties such as curvature or volume growth of the geodesic balls (see, for instance, \cite{LT} and \cite{V}). Thus, in this article we will be able to study the Brownian motion on complete spacelike hypersurfaces once we obtain the geometric consequences that the holographic principle has on these Riemannian manifolds. 

\section{Spacelike hypersurfaces in general ambient spacetimes}
\label{s2} 

Once we have established the formulation of the holographic principle that will be used along this paper as well as the relation between parabolicity of a Riemannian manifold and recurrence of the Brownian motion, in this section we will obtain several results concerning the geometric restrictions that the holographic principle imposes on complete (non-compact) spacelike hypersurfaces immersed in general ambient spacetimes of arbitrary dimension. In particular, we have the following result.

\begin{teor}
\label{teobr}
Let $\psi:M \longrightarrow \overline{M}$ be a complete spacelike hypersurface with non-negative Ricci curvature in a spacetime that satisfies the spacelike entropy bound. If the entropy density is bounded from below by a positive constant $\sigma_0$, then the Brownian motion on $M$ is transient. In particular, if $M$ is a simply connected spacelike surface, it is conformally equivalent to $\mathbb{H}^2$.
\end{teor}

\begin{demo}
Let us suppose that the spacetime satisfies the spacelike entropy bound. If the entropy density is bounded from below by a constant $\sigma_0 > 0$, we know that the geodesic balls in $M$ satisfy

\begin{equation}
\label{volare}
\sigma_0 \mathrm{Vol}(B_R) \leq S(B_R)  \leq \frac{\mathrm{Area}(\partial B_R)}{4}.
\end{equation}

\noindent Hence, using the coarea formula in (\ref{volare}) we obtain

$$ 4 \sigma_0 \leq \frac{d}{d R} \log (\mathrm{Vol}(B_R)).$$

\noindent Integrating this, we get

\begin{equation}
\label{volbr}
\mathrm{Vol}(B_R) \geq e^{4 \sigma_0 R}.
\end{equation}

\noindent Thus, we have

\begin{equation}
\label{intrvol}
\int_1^{+\infty} \frac{R \ dR}{\mathrm{Vol}(B_R)} \leq \int_1^{+\infty} \frac{R \ dR}{e^{4 \sigma_0 R}} < + \infty.
\end{equation}

It was proved in \cite{V} that for Riemannian manifolds of non-negative Ricci curvature the finiteness of the previous integral is equivalent to the transience of the Brownian motion. The last statement of the theorem follows from \cite[Thm. 5.1]{G}.
\end{demo}

\begin{rem}
\normalfont
Note that the converse does not hold in general. For instance, let us consider the spacelike hyperplanes $\{t_0\}\times \mathbb{R}^3, \ t_0 \in \mathbb{R}$, in the Lorentz-Minkowski spacetime $\mathbb{L}^4$. Clearly, the Brownian motion is transient on these spacelike hyperplanes. However, as it was shown in \cite{FS} we obtain a violation of the spacelike entropy bound by considering a constant entropy density $\sigma_0$ in $\{t_0\}\times \mathbb{R}^3, \ t_0 \in \mathbb{R}$ and a geodesic ball of radius $R > \frac{3}{4 \sigma_0}$.
\end{rem}

Indeed, we can extend our result to the case of spacelike hypersurfaces with negative Ricci curvature with a controlled radial decay immersed in a spacetime that obeys the spacelike entropy bound with a more general entropy distribution. Namely, if the inverse of the entropy distribution is in $L^1([0, + \infty))$ (from now on, we will denote this function space by $L^1(+\infty)$) we obtain the next theorem.

\begin{teor}
\label{teoli}
Let $\psi:M \longrightarrow \overline{M}$ be a complete spacelike hypersurface in a spacetime that satisfies the spacelike entropy bound for an entropy distribution that satisfies

\begin{equation}
\label{entl2}
\frac{1}{S(B_R)} \in L^1(+\infty).
\end{equation}

\noindent Assume that there exists a constant $C_1 > 0$ such that the Ricci curvature of $M$ satisfies

\begin{equation}
\label{ricrad}
\mathrm{Ric}(p) \geq -C_1 \rho^{-2}(p),
\end{equation} 

\noindent for all $p \in M$, being $\rho(p)$ the distance function to a fixed point $x \in M$. If there exist $q \in M$ and $C_2 > 0$ such that the volume comparison condition

\begin{equation}
\label{volcomp}
\mathrm{Vol}(B_R (q)) \leq C_2 \mathrm{Vol}(B_{\frac{R}{2}} (p)), 
\end{equation}

\noindent holds for all $p \in \partial B_R (q)$, then the Brownian motion on $M$ is transient.
\end{teor}

\begin{demo}
From the spacelike entropy bound, we obtain that (\ref{entl2}) yields to

\begin{equation}
\label{voll11}
\frac{1}{\mathrm{Area}(\partial B_R)} \in L^1(+\infty).
\end{equation}

Now, it is an easy calculus exercise to show that (\ref{voll11}) implies the next volume growth condition (see \cite[Lemma 5.10]{BMR1}).

\begin{equation}
\label{voll2}
\frac{R}{\mathrm{Vol}(B_R)} \in L^1(+\infty).
\end{equation}

By means of \cite[Thm. 1.9]{LT} we obtain that under our assumptions (\ref{voll2}) is equivalent to the non-parabolicity of $M$.
\end{demo}

\begin{rem}
\normalfont
Since the property of existence of positive Green's functions is quasi-isometric invariant (see \cite[Cor. 5.3]{G}), we can also prove the transience of the Brownian motion on a spacelike hypersurface $M$ in these ambient spacetimes if $M$ is quasi-isometric to a manifold satisfying the assumptions of Theorem \ref{teoli}. In fact, although the volume comparison condition (\ref{volcomp}) may seem restrictive, it is actually satisfied on any complete non-compact Riemannian manifold with non-negative Ricci curvature due to the Bishop comparison theorem \cite{BC}.
\end{rem}

Moreover, the validity of the holographic principle for an entropy distribution satisfying (\ref{entl2}) also implies the transience of the Brownian motion in the particular case where the spacelike hypersurfaces are model manifolds in the sense of Greene and Wu \cite[Def. 2.11]{GW}. Namely, we have

\begin{teor}
\label{teomodel}
Let $\psi: M \longrightarrow \overline{M}$ be a spacelike hypersurface which is a model manifold in a spacetime $\overline{M}$ that satisfies the spacelike entropy bound for an entropy distribution that satisfies (\ref{entl2}). Then, the Brownian motion on $M$ is transient.
\end{teor}

\begin{demo}
If the spacetime satisfies the spacelike entropy bound, from (\ref{entl2}) we can deduce

\begin{equation}
\label{voll1}
\frac{1}{\mathrm{Area}(\partial B_R)} \in L^1(+\infty).
\end{equation}

The result follows from the fact that for a model manifold $M$ condition (\ref{voll1}) is equivalent to non-parabolicity (see \cite[Prop. 1.1]{I}).
\end{demo}

By means of these results we can obtain spatially open spacetimes where the spacelike entropy bound does not hold. In particular, the spacelike entropy bound is violated in those spacetimes that admit a parabolic spacelike hypersurface, known as spatially parabolic spacetimes \cite{RRS}. Indeed, from Theorem \ref{teobr} we have the following corollary.

\begin{coro}
\label{coropa1}
Let $\psi:M \longrightarrow \overline{M}$ be a parabolic spacelike hypersurface with non-negative Ricci curvature. Then, the spacelike entropy bound does not hold on $M$ for an entropy density which is bounded from below by a positive constant.
\end{coro}

\begin{demo}
If $M$ satisfies the spacelike entropy bound for an entropy density $\sigma \geq \sigma_0 >0$, Theorem \ref{teobr} yields to the non-parabolicity of $M$, reaching a contradiction.
\end{demo}

\section{Spacelike hypersurfaces in GRW spacetimes}
\label{segrw}

In this section we will obtain more geometric consequences of the holographic principle in the family of models known as Generalized Robertson-Walker (GRW) spacetimes. By a GRW spacetime we mean a product manifold $\overline{M} = I \times F$ of an open interval $I$ of the real line $\mathbb{R}$ and an $n(\geq 2)$-dimensional (connected) Riemannian manifold $(F,g_F)$, endowed with the Lorentzian metric

$$\overline{g} = -\pi^*_{I} (dt^2) +f(\pi_{I})^2 \, \pi_{F}^* (g_F),$$

\noindent where $\pi_{I}$ and $\pi_{F}$ denote the projections onto $I$ and
$F$, respectively. 

The Lorentzian manifold $(\overline{M}, \overline{g})$ is a warped product (in the sense of \cite[Chap. 7]{O'N}) with base $(I,-dt^2)$, fiber $(F,g_F)$ and warping function $f$. These spacetimes were introduced in \cite{A-R-S1} to extend the classical notion of Robertson-Walker spacetime to the case where the fiber does not necessarily have constant sectional curvature and has arbitrary dimension. Therefore, they include some well-known spacetimes such as Lorentz-Minkowski spacetime, de Sitter spacetime and Einstein-de Sitter spacetime.

For a spacelike hypersurface $\psi:M \longrightarrow \overline{M}$ in a GRW spacetime, we will denote by $\tau:=\pi_I\circ \psi$ the restriction of $\pi_I$ along $\psi$. Moreover, the time-orientation of $\overline{M}$ allows to take a unique unitary
future-pointing vector field $N \in \mathfrak{X}^\bot(M)$ globally defined
on $M$. If we denote by $A$ the shape operator associated to $N$, the corresponding mean curvature function is given by $H:= -\frac{1}{n} \mathrm{trace}(A)$.

In these ambient spacetimes, reasoning as in \cite[Lemma 3]{PRR2}, we can obtain the following bound for the Ricci curvature of spacelike hypersurfaces. 

\begin{lema}
\label{lemaric}
Let $\psi:M \longrightarrow \overline{M}$ be an $n$-dimensional spacelike hypersurface in a GRW spacetime whose warping function satisfies $(\log f)'' \leq 0$ and whose fiber has non-negative sectional curvature. If the mean curvature function of $M$ satisfies

\begin{equation}
\label{meaf}
H^2 \leq \frac{4 (n-1)}{n^2} \frac{f'(\tau)^2}{f(\tau)^2},
\end{equation}

\noindent then, the Ricci curvature of $M$ is non-negative.
\end{lema}

\begin{demo}
Given $p\in M$, let us take a local orthonormal frame $\left\{E_1,\ldots,E_n \right\}$ around $p$. From the Gauss equation we get that the Ricci curvature of $M$, $\mathrm{Ric}$, satisfies

\begin{eqnarray}
\label{riccyy}
\mathrm{Ric}(Y,Y) &=& \sum_{i=1}^n \overline{g}(\overline{\mathrm{R}}(Y,E_i)E_i,Y) + n H g(AY, Y) + g(A^2 Y, Y) \nonumber \\
&=& \sum_{i=1}^n \overline{g}(\overline{\mathrm{R}}(Y,E_i)E_i,Y) + \left|AY + \frac{n H}{2} Y\right|^2 - \frac{n^2 H^2}{4} |Y|^2,
\end{eqnarray}

\noindent for all $Y \in \mathfrak{X}(M)$. Now, from \cite[Prop. 7.42]{O'N}, we have

\begin{eqnarray}
\label{ricbound}
\sum_{i=1}^n \overline{g}(\overline{\mathrm{R}}(Y,E_i)E_i,Y) &=& \sum_{i=1}^n \overline{g}(\overline{\mathrm{R}}(Y^F,E_i^F)E_i^F,Y^F) +  (n-1) \frac{f'(\tau)^2}{f(\tau)^2} |Y|^2 \nonumber \\
& & -(n-2)(\log f)''(\tau) \, g(Y,\nabla \tau)^2 \nonumber \\
& & - (\log f)''(\tau) |\nabla \tau|^2 |Y|^2,
\end{eqnarray}

\noindent where $Y^F$ and $E_i^F$ denote the projections on the fiber $F$ of $Y$ and $E_i$, respectively, and $\nabla \tau$ is the gradient of $\tau$ on $M$. Therefore, introducing (\ref{ricbound}) in (\ref{riccyy}) and taking our assumptions into account we obtain that the Ricci curvature of $M$ is non-negative.
\end{demo}

\begin{rem}
\normalfont
The assumptions made in this lemma on the GRW spacetime of having a fiber with non-negative sectional curvature and satisfying $(\log f)'' \leq 0$ imply that the Ricci tensor of the spacetime satisfies 

$$\overline{\mathrm{Ric}}(z, z) \geq 0,$$

\noindent for all lightlike vectors $z \in T\overline{M}$. This is known as the null convergence condition, which is an algebraic consequence of the Einstein's field equation closely related with the weak energy condition \cite[Sec. 4.3]{HE}.
\end{rem}

We are now in a position to obtain the next result for spacelike hypersurfaces in spatially open GRW spacetimes that satisfy the spacelike entropy bound.

\begin{teor}
\label{teogrw}
Let $\psi:M \longrightarrow \overline{M}$ be an $n(\geq 3)$-dimensional complete spacelike hypersurface in a GRW spacetime whose warping function satisfies $(\log f)'' \leq 0$ and whose fiber has non-negative sectional curvature. If $\overline{M}$ satisfies the spacelike entropy bound for an entropy density which is bounded from below by a positive constant and the mean curvature function of $M$ satisfies (\ref{meaf}), then the Brownian motion on $M$ is transient.
\end{teor}

\begin{demo}
Under our assumptions, we can use Lemma \ref{lemaric} to ensure that the Ricci curvature of $M$ is non-negative. This and the fact that the spacelike entropy bound holds for an entropy density $\sigma \geq \sigma_0 > 0$ allow us to apply Theorem \ref{teobr} to deduce the transience of the Brownian motion on $M$.
\end{demo}

Note that in Theorem \ref{teogrw} we have omitted the $2$-dimensional case. This is due to the fact that Lemma \ref{lemaric} enables us to find $3$-dimensional GRW spacetimes that violate the spacelike entropy bound. Namely, we get

\begin{prop}
\label{propa}
Let $\psi:M \longrightarrow \overline{M}$ be a complete spacelike surface in a $3$-dimensional GRW spacetime whose warping function satisfies $(\log f)'' \leq 0$ and whose fiber has non-negative sectional curvature. If the mean curvature function of $M$ satisfies

\begin{equation}
\label{mea2}
H^2 \leq \frac{f'(\tau)^2}{f(\tau)^2},
\end{equation}

\noindent then, the spacelike entropy bound for an entropy density $\sigma \geq \sigma_0 >0$ does not hold on $M$.
\end{prop}

\begin{demo}
From Lemma \ref{lemaric} we obtain that the Gaussian curvature of $M$ is non-negative. Since $M$ is complete, a classical result by Ahlfors and Blanc-Fiala-Huber for Riemannian surfaces ensures its parabolicity (see \cite{H}). If the spacelike entropy bound for an entropy density $\sigma \geq \sigma_0 >0$ holded on $M$, Theorem \ref{teobr} would yield to the non-parabolicity of $M$, reaching a contradiction.
\end{demo}

\begin{rem}
\normalfont
Notice that inequalities like (\ref{meaf}) and (\ref{mea2}) that relate the mean curvature function of a spacelike hypersurface with the Hubble function $f'/f$ have been previously used to characterize the spacelike slices in GRW spacetimes (see \cite{PRR3} and references therein). Therefore, in this article we find new applications for this type of inequalities.
\end{rem}

\section*{Acknowledgements} 

The author would like to thank the anonymous referees for their deep reading and valuable suggestions to improve this article. The author is supported by Spanish MINECO and ERDF project MTM2016-78807-C2-1-P.

\end{document}